\documentstyle[epsfig,psfig]{texas}
\begin{document}
\title{New Developments in the
Search for the Topology of the Universe} 
\author{Jean-Philippe Uzan$^1$, Roland Lehoucq$^2$ and
Jean-Pierre Luminet$^3$\\}
\address{(1) D\'epartement de Physique Th\'eorique, Universit\'e de Gen\`eve,\\
24 quai E. Ansermet, CH-1211 Geneva (Switzerland)\\}
\address{(2) CE-Saclay, DSM/DAPNIA/Service d'Astrophysique,\\
F-91191 Gif sur Yvette cedex (France)\\}
\address{(3) D\'epartement d'Astrophysique Relativiste et de Cosmologie,\\
Observatoire de Paris, UPR 176, CNRS, F-92195 Meudon (France).\\}
{\rm Email: uzan@amorgos.unige.ch, roller@discovery.saclay.cea.fr,
jean-pierre.luminet@obspm.fr}
%%%%%%%%%%%%%%%%%%%%%%%%%%%%%%%%%%%%%%%%%%%%%%%%%%%%%%%%%%%%%%%%%%

\begin{abstract}
Multi--connected Universe models with space idenfication scales
smaller than the size of the observable universe produce topological
images in the catalogs of cosmic sources.  In this review, we present
the recent developments for the search of the topology of the universe
focusing on three dimensional methods.  We present the
crystallographic method, we give a new lower bound on the size of
locally Euclidean multi--connected universe model of
$3000\,h^{-1}\,\hbox{Mpc}$ based on this method and a quasar catalog, we
discuss its successes and failures, and the attemps to generalise it.
We finally introduce a new statistical method based on a collecting
correlated pair (CCP) technique.
\end{abstract}

\section{BASICS}

\subsection{Historical elements}

In the standard cosmological framework, the universe is decribed by a
{Friedmann-Lema{i}tre solution}, the spatial sections of which
being usually assumed to be simply--connected.  Einstein equations
being local, they allow us to determine the local geometry but they
give no complete information about the global structure of the
universe, i.e.  about its topology, even if the geometry constrains to
some extent the topology.  This was pointed out by Friedmann himself
just after he proposed its cosmological solution \cite{friedmann23}.
The indeterminacy about the global topology of the cosmological
solutions was raised out as soon as Einstein proposed the first
cosmological solution of his equations~; the {Einstein static
universe} assumed spatial sections with the topology of the
{hypersphere} ${S^3}$ \cite{einstein17}, although de Sitter stressed
that the same geometry could admit as well the projective space
${P^3=S^3/Z^2}$ as spatial sections \cite{desitter17}.  {These two
solutions were locally identical (same metric) but differed by their
topology, i.e.  by the choice of boundary conditions.}

Many arguments were then advanced in favour of a simply--connected
universe, such as the {simplicity} or {economy} principle stating that
one should introduce as few parameters as possible in physical
modelling.  Indeed, such an argument is very unclear.  There has been
some tendency to favor spaces with finite volume.  The eternal and
infinite space of Newtonian physics was for instance leading to
logical difficulties such as the {Olbers paradox} \cite{olber}, and
the finite cosmological solution proposed by Einstein in 1917 was
received as a smart way of solving such paradoxes.  The {Mach's
principle} \cite{mach1,mach2} based on the idea that the local inertia
is determined by the distribution of masses in the whole universe,
also tends to favor universes with a finite volume.  If space is
simply--connected then it is finite if and only if it is locally
elliptic, whereas if multi--connected it can also be locally
hyperbolic or Euclidean.  It has also been argued that infinite spaces
were {unaesthetic} since ``all phenomenon with a non vanishing
probability must happen somewhere else'' \cite{epicure, ellis71}.
This argument was used by Ellis to conclude that if space was infinite
then one can avoid the former conclusion by dropping the assumption of
spatial homogeneity and by arguing that we were living in a particular
place of the universe \cite{ellis71}.  To finish, many arguments
coming from some ideas in {quantum cosmology} are in favor of a finite
volume space \cite{wheeler57,aktaz82,hawking84,gibbons98}.  Indeed,
none of these arguments can help us to determine the shape and size of
our universe.  So, we are lead to the question, {can we detect or
constrain observationnally the topology of our universe~?}

Astonishingly enough, our century has on one hand seen the birth of
the geometrical description of the universe and of a non static curved
spacetime.  On the other, mathematicians have developped the
classification of 3D--manifolds.  The first motivation came from
crystallography, which lead to the classification of Euclidean
3D--manifolds \cite{feodoroff85,bierbach11} and was achieved in 1934
\cite{novacki34}.  The classification of locally elliptic 3-manifold
was set by F. Klein \cite{klein90} and W. Killing \cite{killing91} and
solved by J.A. Wolf in 1960 \cite{wolf60}.  The classification of
locally hyperbolic manifold was started in the 70's by W. Thurston
\cite{thurston70,thurston84} and is not yet achieved
\cite{weeks85b,thurston97,thurston98}.  Thus both elements necessary
to answer the question of cosmic topology were developped at the same
time without interacting so much.  For more details concerning this
interesting modern quest, one can see
\cite{luminet98,luminet99,luminetcosmo,luminetnew}.

\subsection{\bf Mathematical elements}

In relativistic cosmology, our universe is described by a {globally
hyperbolic} 4--manifold {${\cal M}=\Sigma\times R$}
\cite{geroch71,hawking73}, where the spatial sections {$\Sigma$} are
homogeneous and isotropic Riemannian 3--manifolds.  From a topological
point of view, it is convenient to {decribe} such a manifold by its
{fundamental polyhedron} (hereafter FP), which is convex.  Its faces
are associated by pairs through the elements of a {holonomy group}
${\Gamma}$ which is acting freely and discontinuously on $\Sigma$ (see
\cite{wolf83,beardon83,nakahara90} for mathematical definitions and
\cite{luminet95,uzan97,uzan98} for an introduction to topology in the
cosmological context).  The holonomy group is isomorphic to the
fundamental group $\pi_{1}(\Sigma)$.  Using the property (see e.g.
\cite{nakahara90}):
\begin{equation}
\pi_1({\cal M})=\pi_1(\Sigma\times
R)\sim\pi_1(R)\times\pi_1(\Sigma)\sim \pi_1(\Sigma),
\end{equation}
 we deduce that the study of the topology of the universe reduces to
 the study of the topology of its spatial sections.

What are the allowed homogeneous 3--manifolds usable in cosmology~?
According to the sign of the spatial curvature $K$, the universal
covering space of the spatial sections (which will be referred to as
$\widetilde{\Sigma}$) can be described by the {Euclidean space}
{$E^3$}, the {hypersphere} {$S^3$} or the {3-hyperboloid} {$H^3$} if
$K$ vanishes, is positive or negative respectively.  Thus, the
homogeneous and isotropic 3-manifolds will be of the form
\begin{equation}
E^3/\Gamma,\quad S^3/\Gamma\quad\hbox{and}\quad H^3/\Gamma.
\end{equation}
Let us just summarize some of the properties of each family and also
introduce {$G$}, the full {isometry group} of $\widetilde{\Sigma}$
keeping the metric invariant.  Indeed $\Gamma$ will be a discrete
sub-group of $G$.

\begin{enumerate}
\item\underline{locally Euclidean 3--manifolds}: The covering space
in the Euclidean space $\widetilde{\Sigma}=E^3$ and the isometry group
is $G={R}^3\times SO(3)$.  The metric can be written under the form
\begin{equation}
ds^2=a^2(\eta)\left\lbrace-d\eta^2+d\chi^2+\chi^2d\Omega\right\rbrace,
\end{equation}
where $\eta$ is the conformal time, $a(\eta)$ the scale factor, $\chi$
the radial coordinate and $d\Omega$ the unit solid angle.  The
generator of $\Gamma$ are the identity, the translations, the
reflexions and the helicoidal motions.

These transformations generate 18 different spaces \cite{wolf60} among
which 17 are multi--connected and correspond to the 17
crystallographic groups \cite{feodoroff85}.  10 spaces are compact and
among them 6 are orientable.  The description of these spaces can be
found in e.g.  \cite{luminet95}.

\item\underline{locally elliptic 3--manifolds}~: The covering space in
the elliptic space $\widetilde{\Sigma}=S^3$ and the isometry group is
$G=SO(4)$.  The metric can be written as
\begin{equation}
ds^2=a^2(\eta)\left\lbrace-d\eta^2+d\chi^2+\sin^2{\chi}d\Omega\right\rbrace.
\end{equation}
With the curvature radius as unit length, the volume of the
hypersphere is
\begin{equation}
\hbox{Vol}({S}^3)=\int_0^\pi4\pi\sin^2{\chi}d\chi=2\pi^2.
\end{equation}
Wolf \cite{wolf60} gives an exhaustive description of the allowed
discrete groups $\Gamma$.  They are the cyclic groups of order $p$
($p\geq2$), the dihedral groups of order $2m$, and the symmetry groups
of the tetrahedron, octaedron and icosahedron.

If we denote $|\Gamma|$ the order of the holonomy group, it is
straigthforward to show that
\begin{equation}
\hbox{Vol}({S}^3/\Gamma)=2\pi^2/|\Gamma|,
\end{equation}
which tells us that the volume of an elliptic manifold is a
topological invariant (degenerate since two different groups with same
order will have same volume).

\item\underline{locally hyperbolic 3--manifolds}: The covering space
in the hyperbolic space $\widetilde{\Sigma}=H^3$ and the isometry
group is $G=PSL(2,{C})\equiv SL(2,{C})/{Z}_2$.  The metric can be
written as
\begin{equation}
ds^2=a^2(\eta)\left\lbrace-d\eta^2+d\chi^2+\sinh^2{\chi}d\Omega\right\rbrace.
\end{equation}

The classification of compact hyperbolic manifolds is not achieved yet
and we will just describe two of them and give some useful properties.
This classification relies on the rigidity theorem
\cite{mostow73,prasad73} stating that the geometry is fixed by the
topology, a consequence of which being that the volume and other
characteristic lengths are topological invariants, so that
\begin{equation}
\pi_1(X)\sim\pi_1(Y)\Longleftrightarrow \hbox{Vol}(X)=\hbox{Vol}(Y).
\end{equation}
It has also been shown that there is a minimal allowed volume
\cite{gabai96}
\begin{equation}
\hbox{Vol}(\Sigma)\geq \hbox{Vol}_{\rm min}=0.166.
\end{equation}
We also define the outside radius $r_+$, the radius of the smallest
geodesic ball that contains the FP, the inside radius $r_-$, the
radius of the biggest geodesic ball contained in the FP, and the
injectivity radius $r_{\rm inj}$, half the length of the shortest
closed geodesic.  $r_+$, $r_-$ and $r_{\rm inj}$ are altogether
topological invariants.

The classification and the description of the known compact hyperbolic
manifolds can be obtained by using the software {\it SnapPea}
\cite{snappea} which provides the FP, the generators of the holomy
group and the characteristic lengths.  Examples of such manifolds can
be found in e.g.  \cite{luminet95}.  Here we mention for a later use a
manifold of special interest, the Weeks space \cite{weeks85}, which is the
smallest known compact hypermabolic manifold.  Its FP has 18 faces and
its geometrical characteristics are
\begin{equation}
\hbox{Vol} = 0.94272, r_{+} = 0.7525, r_{-} = 0.5192, r_{\rm inj} = 0.2923
\end{equation}
\end{enumerate}

\subsection{\bf Physical and observational introduction}

At the time being, two classes of methods to detect and/or constrain
the topology of the spatial sections of our universe have been
proposed and investigated.  They use two classes of cosmological
observations, namely the 2D cosmic microwave background data and the
3D catalogs of discrete sources.

\begin{enumerate}
\item\underline{2D--methods}: The cosmic microwave background is
composed of all the photons emitted during the decoupling period
between matter and radiation at a redshift of $z\sim1100$.  It is
observed as a black body with a temperature of $2.7$ K with small
fluctuations of order $10^{-5}$.  The temperature anisotropies reflect
the small inhomogeneities that ultimately lead to the observed
structures in the universe.

When studying these anisotropies in a multi--connected universe, one
has to take into account the fact that we are in a compact manifold
and thus that the wavelengths must be discretized.  Stevens {\em et
al.} \cite{stevens93} showed that in a cubic hypertorus one should see
a cut--off in the two--point correlation function if $L/R_H\leq0.8$.
This analysis was then generalised to non cubic hypertorus
\cite{oliveira96} and to the six flat manifolds \cite{levin1}.
However, in compact hyperbolic manifolds there are super-curvature
modes \cite{thurston70}, which implies that no such cut--off exists
\cite{inoue99,bond98}.

It has then been realised that the patterns on such an extended
surface must be correlated if the universe is multi-connected.  Levin
{\em et al.} \cite{levin} studied these correlations in the flat
manifolds and Cornish {\em et al.} \cite{cornish97} developped a
topology independent method to look for a topological signature in the
cosmic microwave background map as expected from the MAP and Planck
satellites.

All these studies have been performed assuming that the small initial
inhomogeneities where generated during an inflationary phase.  There
is however another process that could generate such inhomogeneities,
namely topological defects produced during a symmetry breaking phase
transition in the early universe.  Uzan and Peter \cite{uzan97b}
showed that the topology implies a constraint on the topological
defects network from which it was deduced \cite{uzan98b} that there
will be a cut--off in the angular power spectrum of the temperature
anisotropies, the cut--off value depending only on the characteristic
size of the universe and on the cosmological parameters.  Note that
such considerations also provide another method for constraining the
topology, namely the observation of an extended defect at a redshift
$z$ will give a lower bound on the size of the universe.

In conclusion, at the moment 2D--methods constrain flat manifolds to
$L/R_H\leq0.8$, and there is no convincing constraint on compact
hyperbolic manifolds \cite{aurich}.  Some powerful methods can be hoped
to be used when high resolution cosmic microwave background maps are
achieved.

\item\underline{3D--methods}: In the framework of pre--relativistic
cosmology, it was already pointed out by Schwarzschild that if space
is multi--connected, one could see multiple images of astronomical
objects \cite{sch00}, whereas Friedmann \cite{friedmann23} did the
same in the framework of general relativity.  All 3D--methods for
testing the space topology are based on this fact.  We shall develop
their description in the following sections.
\end{enumerate}

In these Proceedings, we describe the developments of the 3D--methods,
from the Schwarzschild initial idea of the existence of multiple
images of the same object, until its statistical implementation under
the form of {cosmic cristallography}.  We then carefully study the
method and explain why, contrary to what was previously thought, it
won't work in locally hyperbolic universes.  We then present attempts
to generalise it and finish by describing a new and promising ``CCP"
method.  This will be examplified by numerical simulations with
depiction of color pictures in order to make the subject as clear as
possible.

\section{\bf FROM THE ORIGINAL IDEA TO THE CRYSTALLOGRAPHIC METHOD}

\subsection{\bf Initial idea}

\begin{figure}
\centering
\epsfig{figure=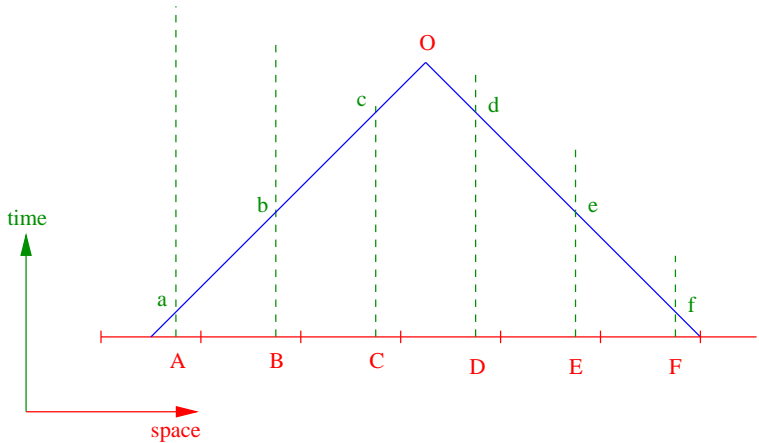,width=10cm}
\caption{a $1+1$ analogy of an multy-connected universe $S^1\times R$.
In the universal covering space, we have plotted the past light cone
of the observer (blue), the trajectories (green) of different
topological images (A-F) of a given object, assuming it is comoving,
and the spacetime point (a-f) where they are observed.  $O$ will see
them at different look-back times.  Indeed in $3+1$ dimension, he will
also see them in different directions in the sky.}
\label{fig4}
\end{figure}

As stressed above, if space is multi--connected, one should see
multiple ``topological" images of the same object.  This is
illustrated on figure \ref{fig4} where we have used a 2--D example.
In the universal covering space, the spacetime trajectories of the
different images ($A,B,\ldots$) of a given object will intersect the
observer past light cone at different times ($a,b,\ldots$).  The same
object will then been seen at different stages of its evolution.
Indeed in a 4D universe, these images will also be seen in different
directions of the sky.

The question is indeed to find methods to implement such a property in
order to detect or to put lower bounds on the size of the universe.

\subsection{\bf Multiple images of individual objects}

Some specific objects have been tried to be recognised individually.
An extensive description of these attempts can be found in
\cite{luminet95} and we just stick to a brief description.

\begin{itemize}
\item Milky Way: Can we recognise our own Galaxy~?  The maximum
distance up to which we would be able to do that has been estimated as
$7.5\, h^{-1}\,\hbox{Mpc}$ by Sokolov and Shvartsmann \cite{sokolov74}
which gave the observational bound $r_+\geq15\, h^{-1}\,\hbox{Mpc}$.
Then, Fagundes and Wichoski \cite{fagundes87} proposed that our galaxy
was a quasar long ago but indeed, for physical reasons due to
realistic quasar models, our Galaxy cannot have been a quasar in its
past.  Fagundes \cite{fagundes89} also studied the occurrence of
images of the Milky Way in the Best hyperbolic space model
\cite{best71}.  Presently, no source has been identified as an image
of our own Galaxy.

\item Galaxy clusters~: Gott used the Coma cluster \cite{gott80} has a
candidate for the search of ghosts images and obtained the bound
$r_+\geq60\, h^{-1}\,\hbox{Mpc}$.  He also performed simulations in
$T^3$ to provide a pattern of clustering and a correlation function in
agreement with observations of nearby galaxies.  Roukema and Edge
\cite{roukema97} proposed 3 candidates as ghost images for the Coma
cluster.  Further investigations are needed.

Sokoloff and Shvartsmann \cite{sokolov74} considered the Abell catalog
of clusters and gave the bound $r_+\geq600\, h^{-1}\,\hbox{Mpc}$ in
Euclidean space, whereas Demianski and Lapucha \cite{demianski87}
searched unsuccessfully for opposite pairs in a catalog of 1889
clusters.  Lehoucq {\em et al.} developped a statistical method that
is presented below to obtain $r_+\geq650\, h^{-1}\,\hbox{Mpc}$ in
Euclidean space and Fetisova {\em et al.} reported a spike about
$125\, h^{-1}\,\hbox{Mpc}$ in the correlation function of rich
clusters, but such an evidence does not necessarily supports
multi--connectedness (see e.g.  \cite{luminet95}).

\item Quasars: Quasars occupy a large volume of the universe
($z\simeq3$).  However, they are probably short-lived compared to the
expected time necessary for a photon to go around the universe.  If
$\tau_q\simeq10^9\,\hbox{years}$, then one can hope to test topology
on scale $\sim r_+\simeq c\tau_q$.  So, they are of interest even if
$\tau_q\simeq10^8\,\hbox{years}$.  Pa\'al \cite{paal} remarked that
although individual quasars may have short lifetime, they may be part
of larger associations which survive longer.

Roukema \cite{roukema96} tried to identify quintuplets and quadruplets
of quasars in a catalog of 4554 quasars.  He found 27 identifications
whereas he found $26\pm1$ identifications in simulations of a
simply--connected universe.  Fagundes \cite{fagundes89} tried to
discuss the redshift contreversy in the Best \cite{best71} model.
\end{itemize}

Until now, the best bound obtained by 3D--methods was $r_+\geq650\,
h^{-1}\,\hbox{Mpc}$ and hold only in Euclidean universe.  There is
still some observational room for the topologies of an hyperbolic
universe, even on scales significantly shorter than the horizon
radius.

All the methods trying to recognize a particular object suffer from
the same major problems:
\begin{enumerate}
\item distance determination: catalogs give the angular position and
the redshift.  One has then to go from redshift space to real space,
which depends on the knowledge of the cosmological constant and of the
density parameter.  This explained why most of the former studies were
performed assuming $\Omega_0=1, \Omega_\Lambda=0$.  This point will be
develop later.  \item peculiar velocities: objects are not comoving,
so that the position of their topological images are shifted.  If
their velocity is of order $v_p\simeq500\,\hbox{km.s}^{-1}$, then,
assuming that the time for a photon to go around the universe can be
estimated by $t\simeq r_+/c$, one obtains the maximal precision on the
determination of position $\delta\chi \simeq (v_p/c)\times r_+$.
\item morphology: when one tries to associate two images to the same
physical object, one has to be aware that, in general, the object will
not be seen in the same angle and at the same stage of its evolution.
It will then be very difficult to recognise it and the identification
will strongly depend on galactic evolution models.
\end{enumerate}

Conversely, the discovery of topology by any means could provide
informations such as peculiar velocities with a better accuracy
\cite{bajtlik}.  We will also be able to see the same object at
different stages of its evolution and on different angles, which will
constrain galactic evolution models.

\subsection{\bf The crystallographic method}

Since there is little chance to recognise different images of a given
object, one can try to detect these images statistically, since for
instance the number of topological images of a single object in a
toroidal universe has been estimated in table \ref{tab} of
\cite{lehoucq96}.

\begin{table}[t]
\begin{center}
\begin{tabular}{*{4}{c}}
\multicolumn{4}{c}{}\\
\multicolumn{4}{c}{\large\bf Number of topological images of a
single object}\\
\multicolumn{4}{c}{}\\
\hline 
$L$ & $\quad\quad N(z<1000$) & $\quad\quad\quad N(z<4)$ &$N(z<1)$\\[0.5ex]
\hline $500$ & 7000 & 1200 & 180 \\[0.5ex]
1500 & 279 & 45 & 7 \\[0.5ex]
2500 & 60 & 10 & 1.5 \\[0.5ex]
\hline
\end{tabular}
\caption{Number of topological images of a
single object in a toroidal universe with characteristic size $L$ (in
$h^{-1}$ Mpc). Reprinted from Lehoucq {\em et al.} \cite{lehoucq96}.}
\label{tab}
\end{center}
\end{table}

The crystallographic method \cite{lehoucq96} is based on a property of
multi--connected universes according to which each topological image
of a given object is linked to each other one by the holonomies of
space.  Indeed, we do not know these holomies as far as we have not
determined the topology, but we know that they are isometries.  For
instance in locally Euclidean universes, to each holonomy is
associated a distance $\lambda$, equal to the length of the
translation by which the fundamental domain is moved to produce the
tessellation in the covering space.  Assuming the FP contains $N$
objects (e.g.  galaxy clusters), if we calculate the mutual
3D--distances between every pair of topological images (inside the
particle horizon), the distances $\lambda$ will occur $N$ times for
each copy of the fundamental domain, and all other distances will be
spread in a smooth way between zero and two times the horizon
distance.  In a histogram plotting the number of pairs versus their 3D
separations, the distances $\lambda$ will thus produce spikes.
Simulations indeed showed that the pairs between two topological
images of the same object drastically emerge from ordinary pairs
\cite{lehoucq96} in the histogram.  The applicability of the method in
Euclidean spaces has also been discussed by Fagundes and Gaussmann
\cite{fagundes97} when the size of the physical space is comparable to
the horizon size.

Two kind of catalogs of astronomical objects can be thought of to
apply this method: the galaxy cluster catalogs, which typically have
a redshift depth $z=1$, and the quasars catalogs, which typically
extend to $z=3$.  Concerning quasars, even if their lifetime is
probably too short to be good candidates for producing topological
images, they are usually part of systems that have a much larger
lifetime \cite{paal}.  Let us stress that a recent survey
\cite{hubble} in the Hubble Deep Field south NICMOS field found 17
galaxies with a redshift between 5 and 10 and 5 galaxies with a
redshift above 10, among a total of 323 galaxies.  This can let us
hope to apply this method to deeper catalogs in the future.  The
angular resolution needed is given by the fact that the objects have a
peculiar velocity and that they will not be seen at exactly the same
position \cite{luminet95}.  Note that the crystallographic method,
contrary to the ``direct'' method which would try to recognize
topological images of individual objects, is not plagued by the fact
that topological images of the same object are seen at different
stages of its evolution.

\begin{figure}
\centering
\epsfig{figure=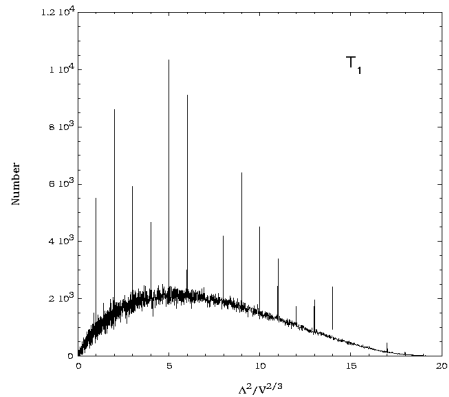,width=6cm}
\caption{Application of the crystallographic method in a cubic torus
universe, from \cite{lehoucq96}.  Here, $\Lambda$ corresponds to the
separation distance and $V$ to the volume of the manifold.}
\label{fig5}
\end{figure}

We present on figure \ref{fig5} an example from \cite{lehoucq96} for
the flat torus with a cubic FP.

\begin{figure}
\centering
\epsfig{figure=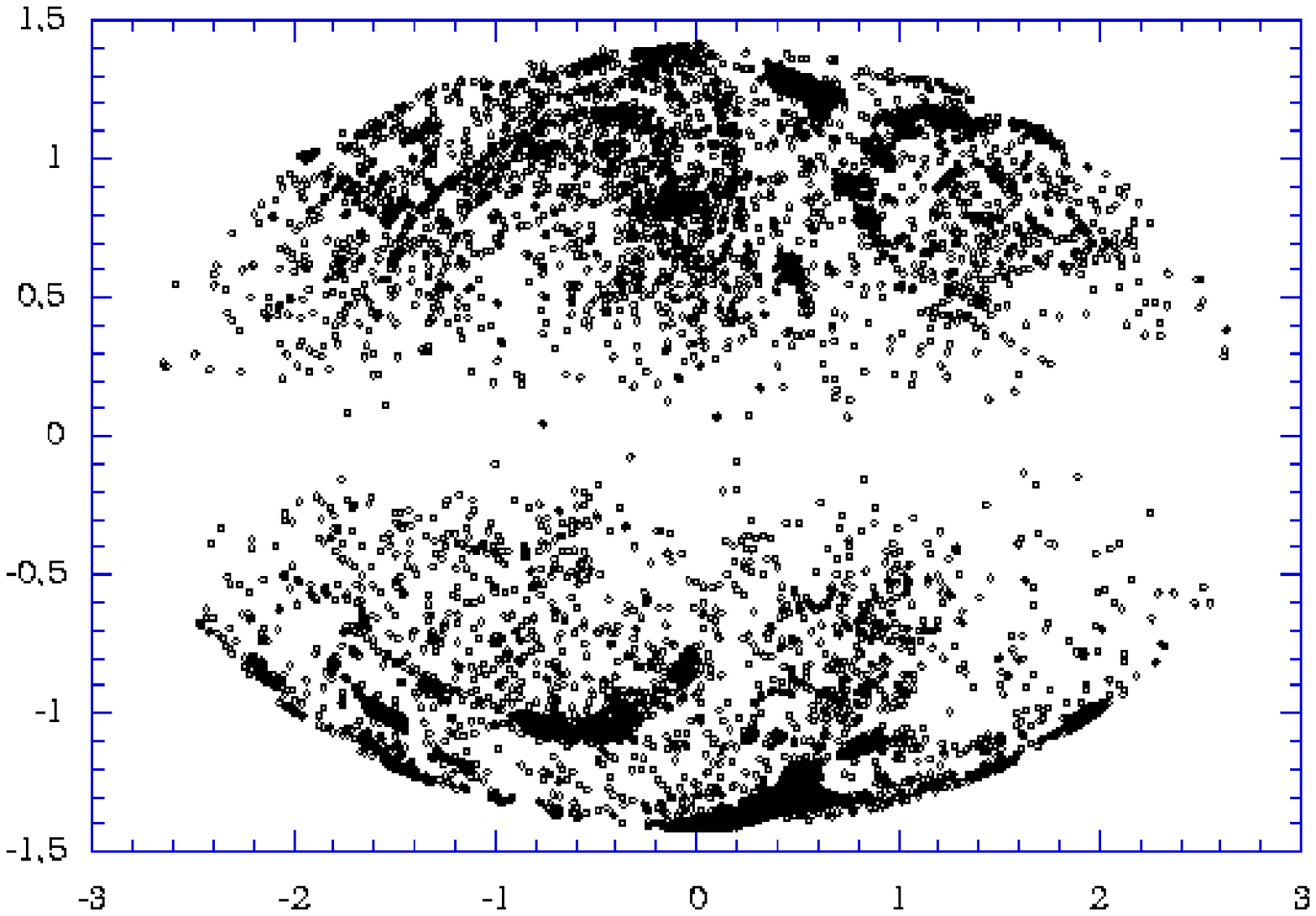,width=6cm}
\epsfig{figure=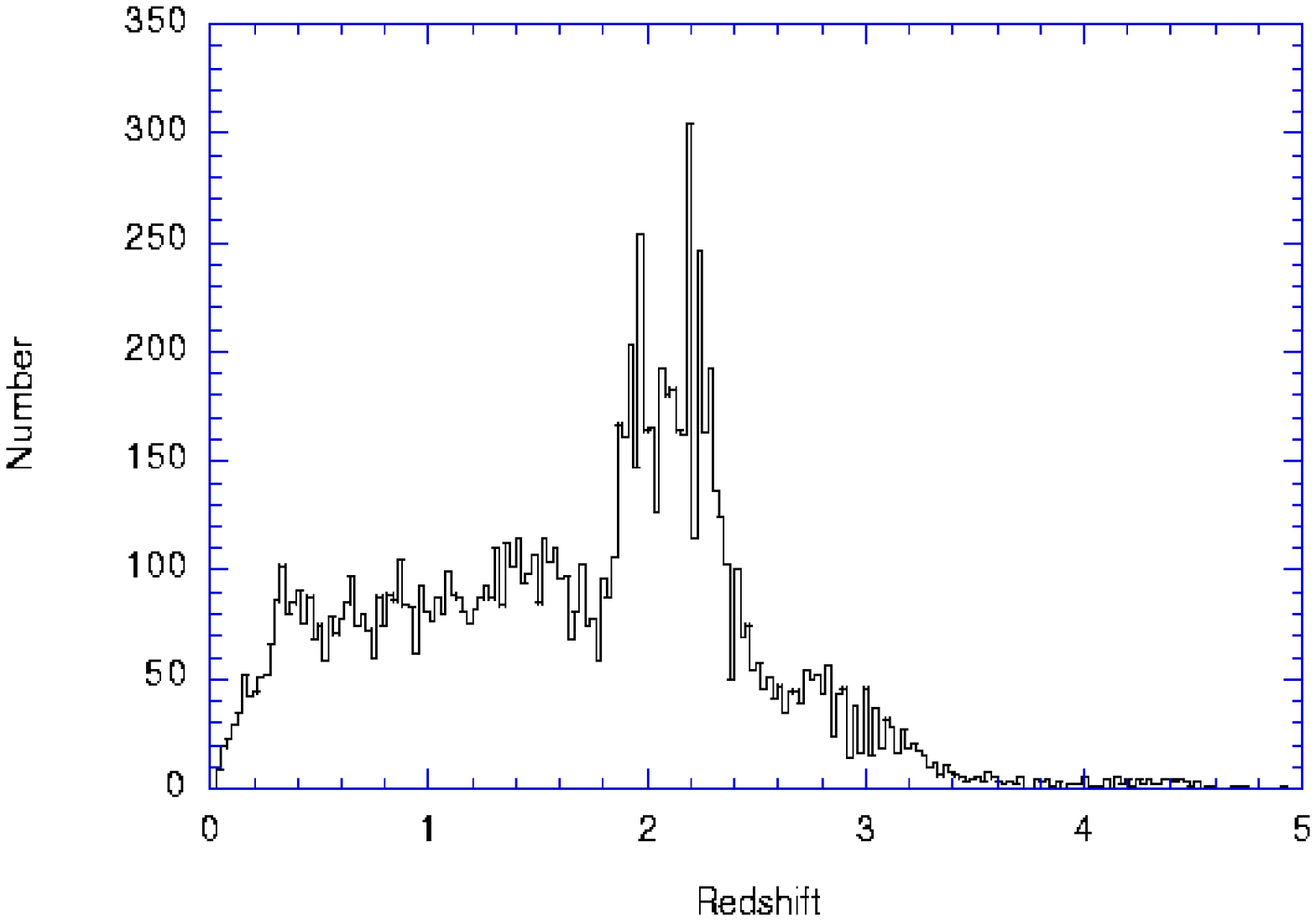,width=6cm} 
\caption{The quasar
catalog: (left) the distribution of objects on the celestial sphere
in Hammer--Aitoff equal area projection and (right) the redshift
distribution.}
\label{catq}
\end{figure}

\begin{figure}
\centering
\epsfig{figure=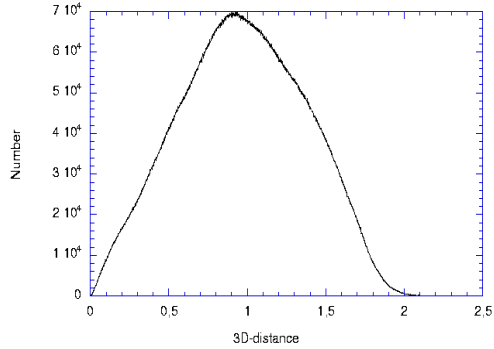,width=7cm}
\caption{The pair separation histogram for the quasar catalog (see
figure \ref{catq}) exhibits no topological signature when assuming a
locally Euclidean geometry.}
\label{psh}
\end{figure}

Now, we apply this method to a catalog \cite{quascat} containing more
than 11,000 quasars up to a redshift of $z\simeq3.25$ (see figure
\ref{catq} where we have depicted its projection on the celestial
sphere and the redshift distribution of its objects).  The pair
separation histogram is plotted on figure \ref{psh} and it exhibits no
topological signature.  Since respectively 60~\%, 80~\% and 95~\% of
the sources are below the redshifts 2, 2.3 and 3, we only test scales
smaller that $z\simeq 3$.

In a locally Euclidean universe, the distance-redshift relation is
given by \cite{peebles}
\begin{equation}
\chi(z)=\frac{2c}{a_0H_0}\left(1-\frac{1}{\sqrt{1+z}}\right).
\end{equation}
We simulated a catalog with the same number of objects and the same
redshift distribution to reproduce the real catalog depicted in figure
\ref{catq}. Varying the size of cubic hypertorus, a spike appears when
$L_0\simeq3000\,h^{-1}\,\hbox{Mpc}$ for the simulated catalog and thus
we obtain the new bound on the size of a locally Euclidean manifold as
\begin{equation}
L_0\geq 3000\,h^{-1}\,\hbox{Mpc},
\end{equation}
which is the best constraint presently available with 3D--methods for
locally Euclidean manifolds.  Note that it corresponds to
\begin{equation}
\frac{L_0}{R_H}\geq0.5,
\end{equation}
which is comparable to the bound obtained with 2D--methods.  Note
that this result has not been published elsewhere.


\section{\bf WHY THE CRYSTALLOGRAPHIC METHOD DOES NOT WORK IN
HYPERBOLIC SPACES}

\begin{figure}
\centering
\epsfig{figure=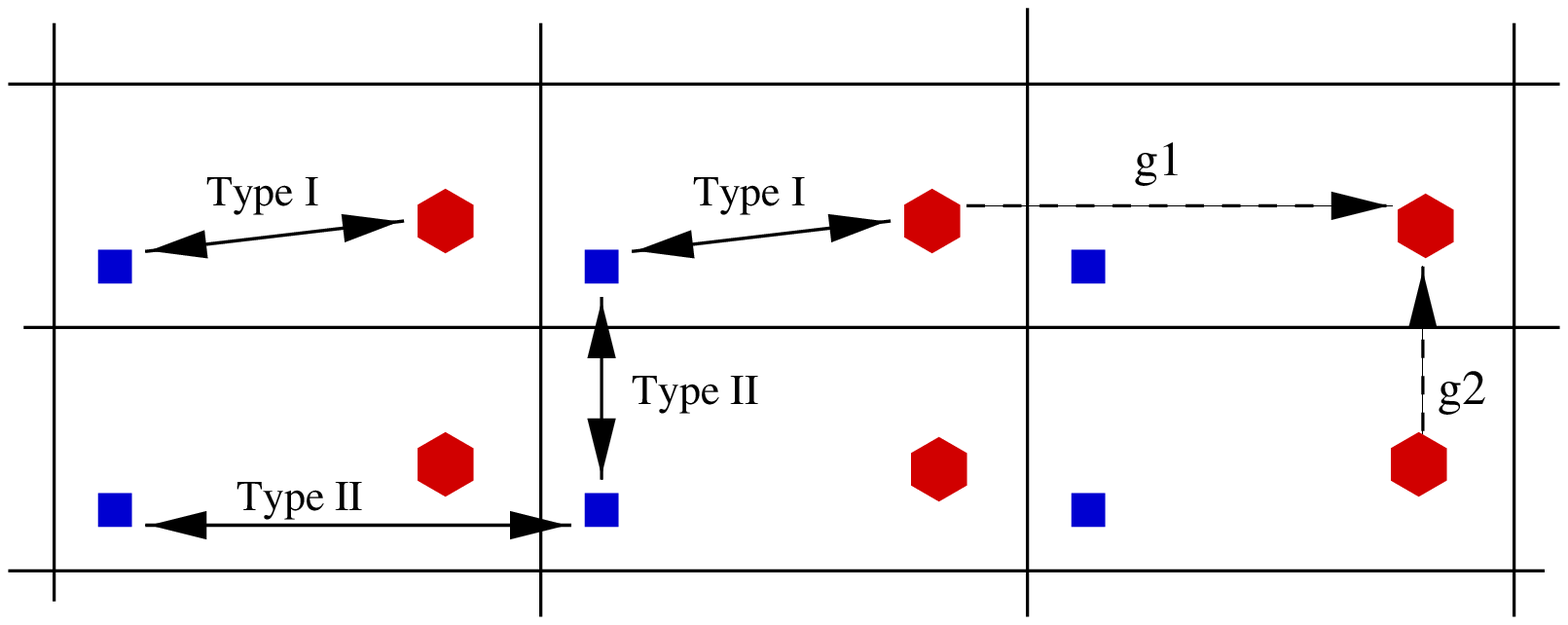,width=8cm}
\caption{Difference between type I and type II pairs in the example of
the two dimensional torus.  The translations $g_1$, $g_2$ $g_1^{-1}$
and $g_2^{-1}$ are the generators of the holonomy group.  Type I pairs
are the ones between the ghosts of two distinct objects (hexagon or
square), and type II pairs are the ones between two topological images
of the same object.}
\label{fig6}
\end{figure}

When one looks carefully at the crystallographic method, one must
wonder about the {origin of the spikes}.  As it has recently been
explained \cite{lehoucq99}, two kinds of pairs can create a spike,
namely (we keep the notation and vocabulary introduced in
\cite{lehoucq99})
\begin{enumerate}
\item {\it Type I pairs} of the form $\lbrace g(x),g(y)\rbrace$, since
$\hbox{dist}[g(x),g(y)]=\hbox{dist}[x,y]$ for all points and all
elements $g$ of $\Gamma$.  \item {\it Type II pairs} of the form
$\lbrace x,g(x)\rbrace$ if $\hbox{dist}[x,g(x)]=\hbox{dist}[y,g(y)]$
for at least some points and elements $g$ of $\Gamma$.
\end{enumerate}
Both families of pairs are depicted on figure \ref{fig6} for the
example of the 2--torus.

\subsection{\bf Clifford translations}

A Clifford translation is an isometry $g$ such that the displacement
function $\hbox{dist}(x,g(x))$ is constant \cite{thurston70} .  This
is precisely what is required to get type II pairs in the histogram.

In compact hyperbolic manifolds, $\hbox{dist}[x,g(x)]$ always depends
on the position $x$ of the source (see \cite{lehoucq99} for details).

In elliptic spaces, distances are position-independent whenever the
holonomy is a Clifford translation.  All finite groups of Clifford
translations of spheres are the cyclic group, the binary dihedral,
tetrahedral, octahedral and icosahedral groups \cite{wolf67}, from
which we deduce \cite{lehoucq99} that the covering transformations
which move a source to its nearest neighbours are Clifford
translations (although the transformations to more distant neighbours
might not be), and type II pairs can be produced.

In locally Euclidean universes, type I and type II pairs are both
present.  The reason is that the 3-torus has the very special property
that the separation distance of gg-pairs (i.e.  any pair of images
comprising an original and one of its ghosts, or two ghosts of the
same object) is independent of the location of the source.  In other
Euclidean spaces the spectrum of gg-pair distances varies with the
location of the source.  However all closed Euclidean 3-manifolds have
the 3-torus as a covering space, so for each such manifold there will
be some distances which are independent of the location of the source.
As a consequence, the topological signal expected in the histogram
from type I and type II pairs clearly stands out, as was shown in the
simulations of \cite{lehoucq96}.\\

In conclusion, type II pairs exist only in elliptic and Euclidean
universes.  The question we still have to answer to judge the
efficiency of the crystallographic method in the general case is:
what about the amplitude of the spikes related to type I pairs only~?

\subsection{\bf Cosmological parameters}

\begin{figure}
\centering
\epsfig{figure=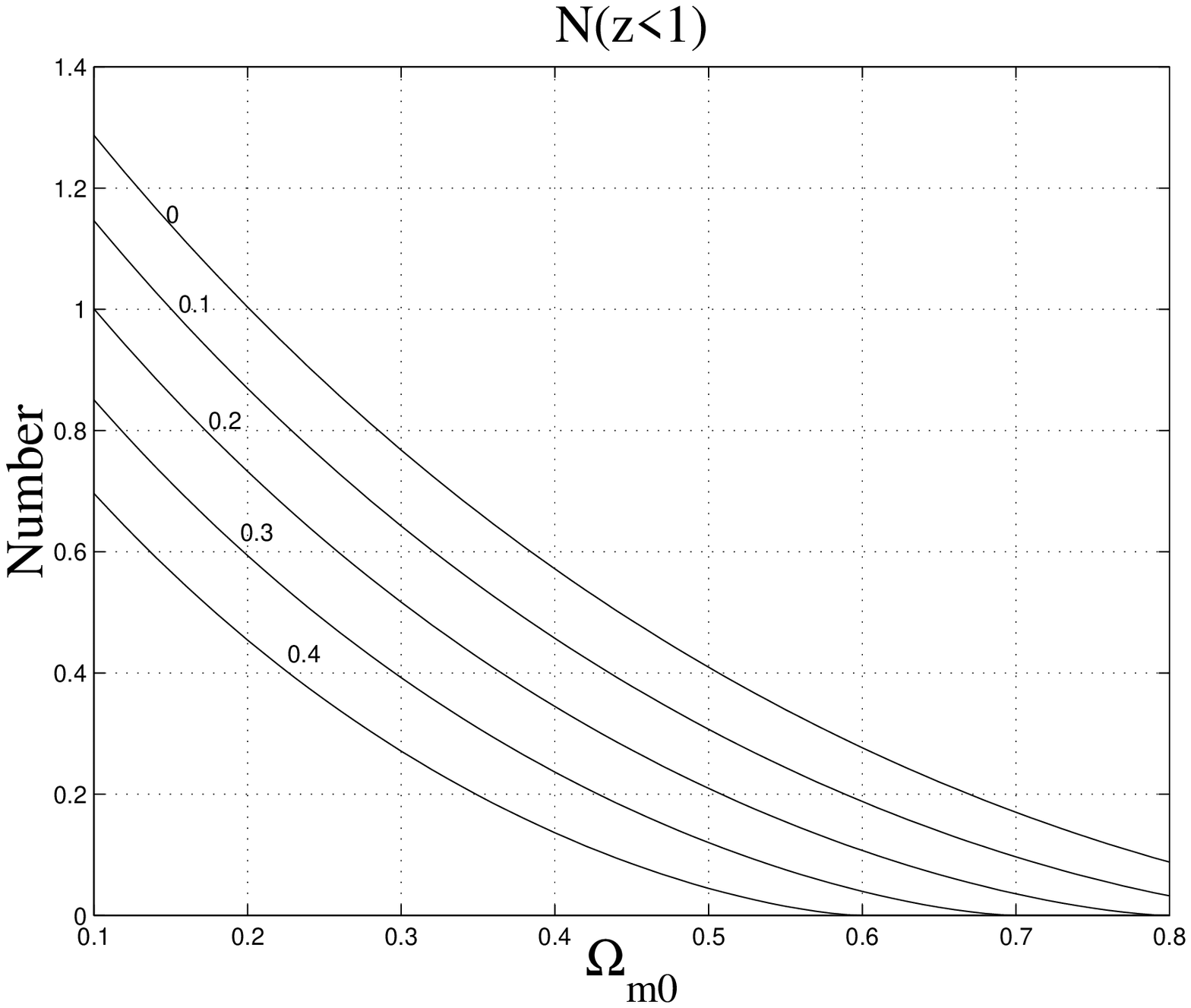,width=6cm}
\epsfig{figure=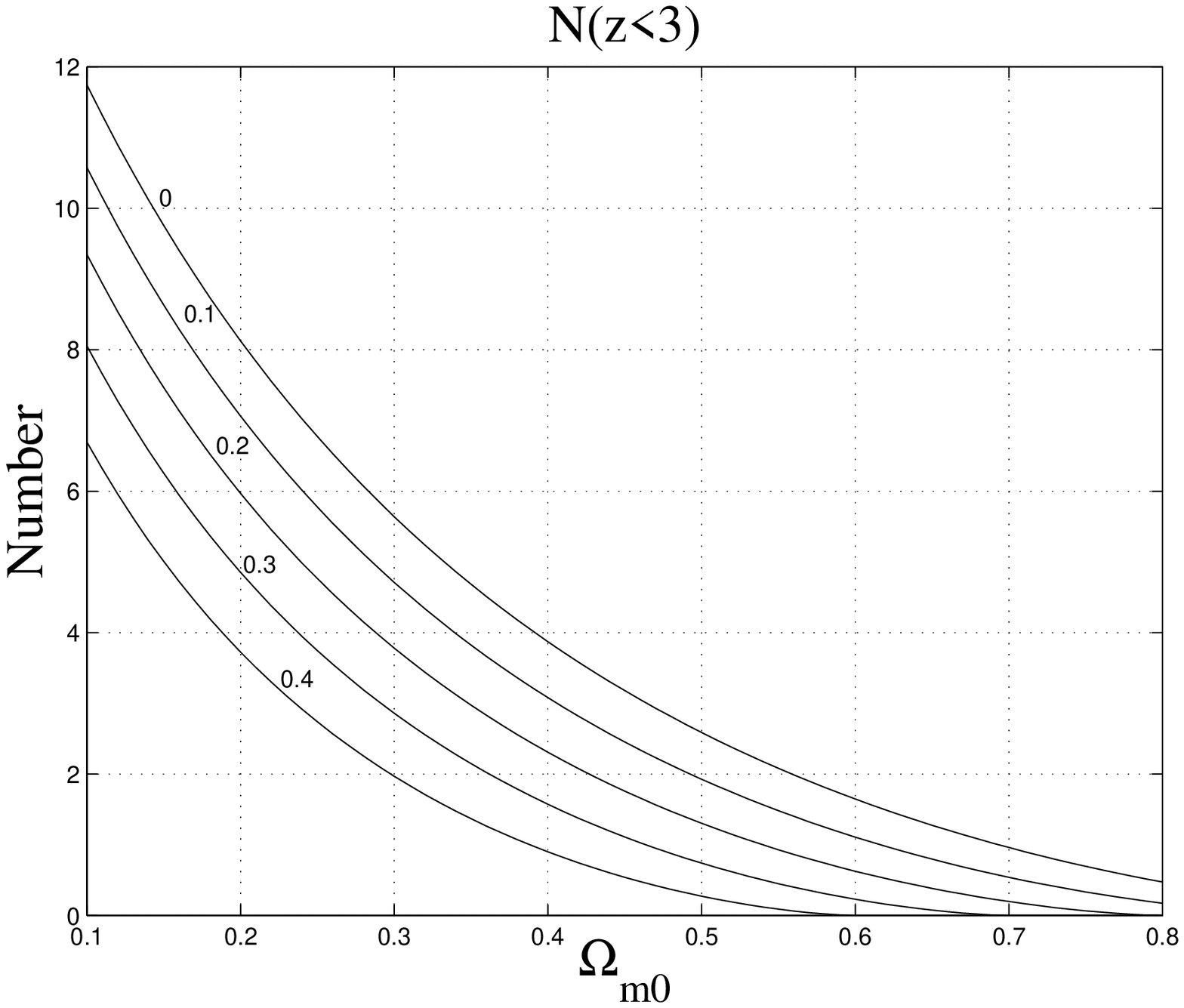,width=6cm} 
\caption{The number of copies of
an object belonging to a catalog of clusters (left) and a catalog of
quasars (right) as a function of the cosmological parameters.}
\label{fig8}
\end{figure}

If the injectivity radius of the space manifold is smaller than the
calatog's length scale, type I pairs will always be present, whatever
the curvature.  Their number roughly equals the number of copies of
the fundamental domain within the catalog's limits.  Following
\cite{lehoucq99}, this number can easily been estimated by computing
the ratio between the volume of the geodesic sphere of radius
$\chi(z)$ and the volume of the manifold.  The calculation (in the
case of hyperbolic universes) leads to
\begin{equation}
N(\Omega_{m0},\Omega_{\Lambda0};z<Z)=\frac{\pi\left(\sinh{2\chi(Z)}
-2\chi(Z)\right)}{\hbox{Vol}(\Sigma)},
\end{equation}
where $\chi(Z)$ is the radial distance corresponding to the redshift
$Z$, given
\begin{equation}
\chi(z)\equiv\int_{a_0}^a\frac{da}{a\dot a}=
\int_{\frac{1}{1+z}}^1\frac{\sqrt{1-\Omega_{m0}-\Omega_{\Lambda0}}dx}
{x\sqrt{\Omega_{\Lambda0}x^2+(1-\Omega_{m0}-\Omega_{\Lambda0})
+\frac{\Omega_{m0}}{x}}},\label{chi1}
\end{equation}
where $\Omega_{m0}$ and $\Omega_{\Lambda}$ are respectively the
density parameter and the cosmological constant.  We have plotted on
figure \ref{fig8} an estimation of the number of copies of an object
taken respectively in a catalog of galaxy clusters and in a catalog of
quasars, as a function of the cosmological parameters.

The small number of type I pairs in hyperbolic manifolds is due to the
property that the volume of the manifold is fixed once the topology is 
determined (the {\it rigidity theorem}), contrarily to Euclidean
spaces where the characteristic sizes and the volume of the FP can be
chosen at will (since $K=0$, the geometry does notimpose any
characteristic size).

In conclusion, type I pairs will indeed contribute to spikes in the
pair separation histogram but the amplitude of these spikes will be of
the same order than the statistical noise.

\subsection{\bf Numerical simulations}

The former conclusion can be tested numerically.  For that purpose, we
concentrate on locally hyperbolic three--dimensional manifolds.  It
can be first embedded in a four--dimensional Minkowski space by
introducing the set of coordinates $(x^\mu)_{\mu=0..3}$ related to the
intrinsic coordinates $(\chi,\theta,\varphi)$ through (see e.g.
\cite{wolf67,coxeter65})
\begin{eqnarray}
x_0&=&\cosh{\chi}\nonumber\\
x_1&=&\sinh{\chi}\sin{\theta}\sin{\varphi}\nonumber\\
x_2&=&\sinh{\chi}\sin{\theta}\cos{\varphi} \nonumber\\
x_3&=&\sinh{\chi}\cos{\theta},
\end{eqnarray}
so that the three--dimensional hyperboloid $H^3$ has the equation
\begin{equation}
-x_0^2+x_1^2+x_2^2+x_3^2=-1.
\end{equation}
With these notations, the comoving spatial distance between two points
of comoving coordinates $x$ and $y$ can be computed directly in the
Minkowski space by \cite{fagundes89}
\begin{equation}
d[x,y]=\arg\cosh{\left[x^\mu y_\mu\right]},
\end{equation}
where $x_\mu=\eta_{\mu\nu}x^\nu$, $\eta_{\mu\nu}$ being the
Minkowskian metric coefficients.

To generate an idealised catalog, ${\cal C}$, we first distribute {\it
homogeneously} $N$ objects in the FP, then we by unfold the catalog by
applying the generators of the holonomy group.  Now, to generate a
catalog with the required depth $z_{\rm max}$ in redshift, the set of
all points obtained as above must be truncated in order to keep only
the points $x$ such that $\hbox{dist}[0,x]\leq
\chi(\Omega_{m0},\Omega_{\Lambda0};z_{\rm max})$, given by equation
(\ref{chi1}).  This amounts to select objects located within the
geodesic ball of radius $\chi(\Omega_{m0},\Omega_{\Lambda0};z_{\rm
max})$ centered onto an observer right at the centre.

\begin{figure}
\centering
\epsfig{figure=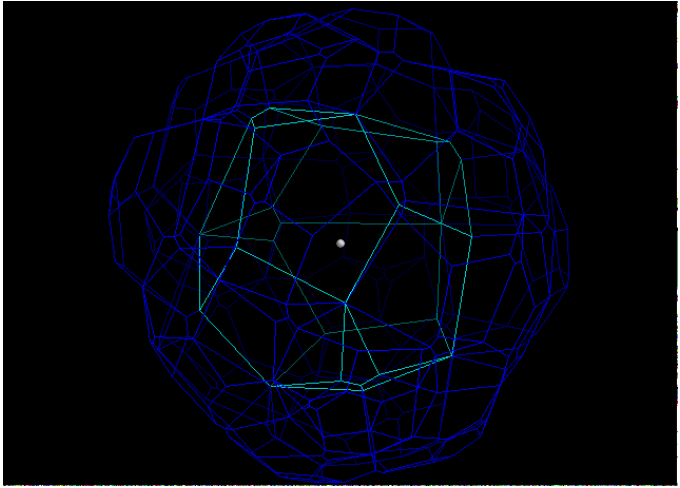,width=9cm} \caption{Unfolding the
Weeks manifold: we have represented the FP and its 18 nearest copies
in Klein coordinates.}
\label{fig9}
\end{figure}

\begin{figure}
\centering
\epsfig{figure=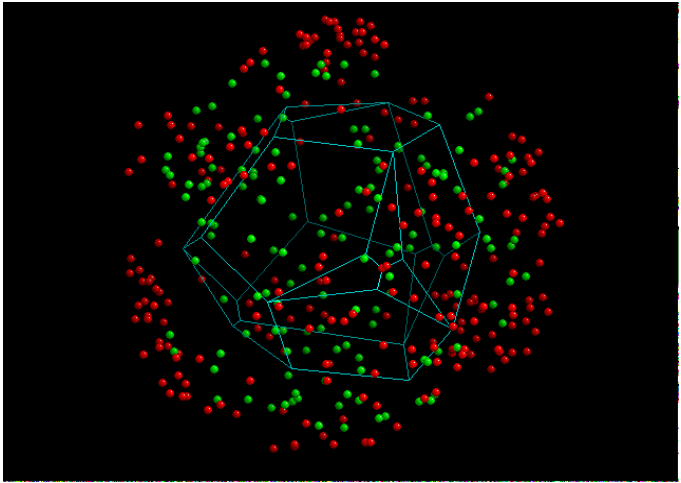,width=9cm}
\caption{Building a catalog ${\cal C}$ for the Weeks manifold.  The
green dots are objects located within the horizon.  The red ones lie
outside and will be rejected from ${\cal C}$.}
\label{fig10}
\end{figure}

The unfolding of Weeks manifold is shown in figure \ref{fig9}, where
we have depicted the central cell and its 18 nearest neighbours in
Klein coordinates.  In figure \ref{fig10}, we have plotted a simulated
catalog showing all points, including those outside the horizon
distance.


We then compute all the 3D separations between all the pairs of ${\cal
C}(z)$ and we plot the histogram of the number of pairs with a given
separation.\\

Indeed the former procedure applies when the observer stands at the
center of the polyhedron $(\chi=0)$.  In \cite{lehoucq99}, we explain
how to take into account an off-centered observer.\\

\begin{figure}
\centering
\epsfig{figure=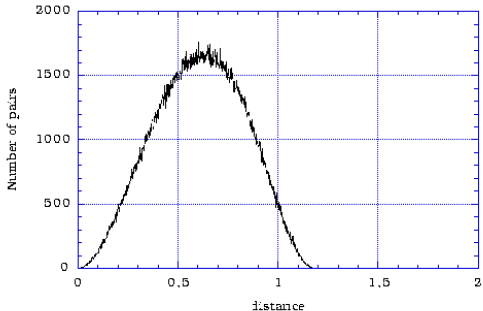,width=7cm}
\caption{Pair separation histogram for a universe with the topology of
the Weeks manifold, $\Omega_0=0.2$, $\Omega_\Lambda=0$, for a
simulated catalog of cosmic objects with depth $z_{\rm max}=1$.  No
spike stands out.}
\label{fig11}
\end{figure}

In Lehoucq et al.  \cite{lehoucq99}, we have generated some pair
histograms in a universe whose spatial sections have the topology of
the Weeks manifold, using $N=1000$ objects in the FP. In figure
\ref{fig11} we give an example wher $\Omega_\Lambda=0$,
$\Omega_{m0}=0.2$ and where the observer stands at the center of the
polyhedron ($\chi=0$).  The dependence on the cosmological parameters,
on the position of the observer and on the catalog depth where studied
in details in \cite{lehoucq99}.

As expected, none of the plots exhibit spikes.  This is due to the
combined effects of the geometry and of the cosmological parameters
values.

To summarize the failure of cosmic crystallography for hyperbolic
manifolds:
\begin{enumerate}
\item hyperbolic manifolds are such that $\hbox{dist}[x,g(x)]$
depends on $x$, so that there is no amplification for the {\it type II
pairs} $\lbrace x,g(x) \rbrace$, whereas
$\hbox{dist}[x,g(x)]=\hbox{dist}[y,g(y)]$ in the Euclidean case.
This suppresses the corresponding spikes.  \item The spikes associated
to the isometries (i.e.  such that $\forall g\in\Gamma,\quad
\hbox{dist}[g(x),g(y)]=\hbox{dist}[x,y]$) must remain.  But, given the
cosmological parameters, we have shown that the number of topological
images is too low to create such spikes associated to {\it type I
pairs}.
\end{enumerate}

\section{\bf GENERALISING THE CRYSTALLOGRAPHIC METHOD}

The next question is: can we generalise the crystallographic method
in order to make it work in any case~?

Two solutions were proposed almost immediately to improve the
crystallographic method and to extract from 3D catalogs the signature
of type I pairs:
\begin{enumerate}
\item Fagundes and Gaussman \cite{fagundes98} substracted the pair
separation histogram for a simulated catalog (with the same number of
objects and the same cosmological parameters) in a simply-connected
universe to the observed pair separation histogram.  The result is ``a
plot with much oscillation on small scale, modulated by a long
wavelength quasi-sinusoidal pattern''.  The statistical relevance of
this signature still has to be investigated.  To illustrate this point
we show on figure \ref{fig12} the result of a simulation where we have
substracted to realisations on a simply-connected
Friedmann-Lema{i}tre universe.  This has to be compared with figure
1 from \cite{fagundes98} and let us think that is method
is not efficient for determining a relevant signature of the 
topology of the universe.

\begin{figure}
\centering
\epsfig{figure=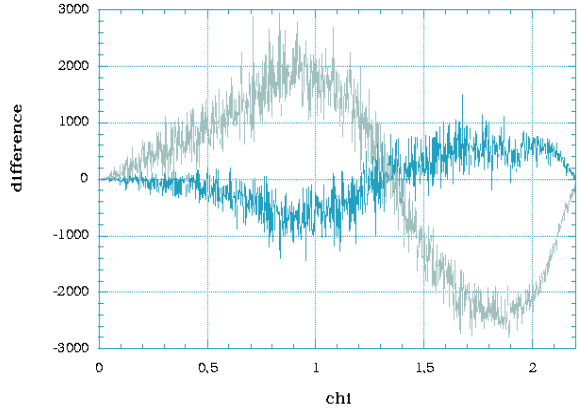,width=8cm}
\caption{The plot for the Weeks manifold [grey] compared to the
difference between two realisations in simply-connected universes with
same local geometry [green].}
\label{fig12}
\end{figure}
\item Gomero {\em et al.} \cite{gomero99} proposed to split the
catalog in ``smaller'' catalogs and average the pair separation
histograms built from each sub--catalog to reduce the statistical
noise and extract a visible signal of the non-translational
isometries.  The feasability of this method has not yet been
demonstrated.
\end{enumerate}

\section{A NEW EFFICIENT STATISTICAL METHOD}

We now present a new method \cite{uzan99} based on the property that
whatever the topology, type I pairs will always exist.  Indeed, as
shown ahead, this does not lead to any observable spike in the pair
separation histogram.  We thus had to find a method to enhance the
topological signal by ``collecting'' all the homologous pairs
together.  This can be achieved by building a {\it collecting
correlated pairs} method (hereafter {\it CCP}).  Such an approach
(described in details in \cite{uzan99} and below) is not able to
determine the exact topology, but will provide a signature of the
compactness of the spatial sections, which is indeed a first step
toward the determination (or the rejection) of the cosmic topology.

In this section, we recall the construction of the CCP--index and the
main resuts we have obtained.

\subsection{\bf Basic idea and general method}

As stressed in the previous section, type I pairs will always exist in
multi--connected spaces as soon as one of the characteristic scales of
the fundamental domain is smaller than the Hubble scale.  Defining
$\left.g_i\right|_{1\leq i\leq 2K}$ as the $2K$ generators of $\Gamma$
and referring to $x$ as the position of the image in the universal
covering space $X$, we have:
\begin{enumerate}
\item $\forall\, x,y\in X$, $\forall g\in\Gamma$,
\begin{eqnarray}
\hbox{dist}[g(x),g(y)]&=&\hbox{dist}[x,y].\label{zero1}
\end{eqnarray}
We will refer to these pairs as $xy$--pairs.  \item $\forall\, x\in
X$, $\forall g_1,g_2\in\Gamma$,
\begin{eqnarray}
\hbox{dist}[g_1(x),g_1\circ
g_2(x)]&=&\hbox{dist}[x,g_2(x)].\label{zero2}
\end{eqnarray}
We will refer to these pairs as $xg(x)$-pairs.
\end{enumerate}
Both the $xy$--pairs and the $xg(x)$-pairs are type I pairs.

To collect all these pairs together and enhance the topological signal,
we define the {\it CCP--index} of a catalog containing $N$ objects as
follows.
\begin{enumerate}
\item We compute all the 3D--distances $\hbox{dist}[x,y]$ for all
points within the catalog's limit.  \item We order all these distances
in a list $\left.d_i\right|_{1\leq i\leq P}$, where $P\equiv N(N-1)/2$
is the number of pairs, such that $d_{i+1}\geq d_i$.  \item We create
a new list of increments defined by
\begin{equation}
\forall i\in [1...P-1],\quad \Delta_i\equiv d_{i+1}-d_i
\end{equation}
 (keeping all the equal distances, if any, in the list).  \item We
 then define the CCP--index ${\cal R}$ as
\begin{equation}
{\cal R} \equiv \frac{{\cal N}}{P-1} \label{CCP},
\end{equation}
where ${\cal N} \equiv \hbox{card}(\lbrace i, \Delta_i=0 \rbrace)$, so
that $0 \leq {\cal R} \leq 1$.

\end{enumerate}
With such a procedure, all type I pairs will contribute to ${\cal N}$.
For instance, if a given distance appears 4 times in the list
$\left.d_i\right|_{1\leq i\leq P}$, it will contribute to 3 counts in
${\cal N}$.  Compared to the old crystallographic method, all the
correlated pairs are gathered into a single spike, instead of being
smoothed out into the noise of the histogram pair separation.

Indeed, in a more realistic situation, one has to take into account
bins of finite width $\epsilon$ and replace ${\cal N}$ by
\begin{eqnarray}
{\cal N}_\epsilon&\equiv& \hbox{card}(\lbrace i,
\Delta_i\in[0,\epsilon[ \rbrace)
\end{eqnarray}
in the computation of ${\cal R}$.  The effect of this ``binning'' is
discussed below.  We now focus on the ``idealised'' version of the
procedure by studying the amplitude of the CCP--index in some
multi-connected models.\\

For that purpose, let us assume that the catalog is obtained from an
initial set of $A$ objects lying in the fundamental domain and that
$B$ copies of the domain are within the catalog's limits ($B=0$ if the
whole observable universe up to the catalog's limit is included inside
a fundamental domain).  The total number of images is $N=A(B+1)$.
Indeed $B$ is usually not an integer but we assume it is, in order to
estimate the amplitude of ${\cal R}$ and compare it with the
result in a simply-connected model.  In \cite{uzan99}, it has been
shown that
\begin{equation}
{\cal N}_{\rm min}=A\left[(A-1)\frac{B}{2}+A \nu_1(\Sigma,B)\right],
\label{amplitude}
\end{equation}
where $\nu_1(\Sigma,B)$ is a function characterising the manifold
$\Sigma$.  It is an increasing function of $B$ which vanishes for
$B=0$.

Indeed, if the holonomy group $\Gamma$ contains Clifford translations
allowing for type II pairs, or if there are ``fake'' pairs (i.e.  such
that $\hbox{dist}[x,y]=\hbox{dist}[u,v]$), ${\cal N}_{\rm min}$ computed
from (\ref{amplitude}) will give a lower bound for the true ${\cal
N}$.

The normalised CCP--index (\ref{CCP}) follows straightforwardly.
${\cal R}$ is a good index for extracting the topological signal since
\begin{enumerate}
\item when $B = 0$ (i.e.  when the fundamental domain is greater than
the catalog's spatial scale), ${\cal R} = 0$, \item when the number of
sources in the fundamental domain becomes large, it behaves as
\begin{equation}
{\cal R} \rightarrow\frac{B+2\nu_1(\Sigma,B)}{(B+1)^2}\quad
\hbox{as}\quad A\rightarrow\infty.
\end{equation}
\end{enumerate}
As shown in \cite{uzan99}, one can compute analytically the CCP--index
in various examples and show that it is a fair indicator of the
existence of at least one compact spatial dimension.

\subsection{\bf Simulations and statistical relevance}

Now that the CCP--index can be computed for any multi-connected space,
we can compare it with its value in a simply-connected
Friedmann-Lema{i}tre model containing the same number objects (i.e.
$A(B+1)$).  ${\cal R}$ is given as a function of $\nu_{1}$, $A$ (the
number of sources in the fundamental domain) and $B$ (the number of
copies of the domain within the observable universe) by:
\begin{equation}
{\cal R} = \frac{A[(2\nu_1-B)A-B]} {(B+1)^2A^2-(B+1)A-2}.
\end{equation}
Indeed, when working with a real catalog, we do not know the radial
distance of cosmic objects but only their redshift.  As seen before,
the determination of the radial distance requires the knowledge of the
cosmological parameters $\Omega_0$ and $\Lambda$ and can be obtained
analytically only when $\Lambda=0$.

Thus, if the universe is multi-connected on sub--horizon scale, the
plot of ${\cal R}$ in terms of $\Omega_0$ and $\Omega_\Lambda$ will
exhibit a spike only when the cosmological parameters have the right
value (as shown on plot 7 of \cite{uzan99}).  If the cosmological
parameters are not exactly known, the distance determination
(\ref{chi1}) will be wrong and the topological signature will be
destroyed (see figure 6 in \cite{uzan99}).

Two consequences follow:
\begin{enumerate}
\item One should span the parameters space $(\Omega_0,\Omega_\Lambda)$
in order to detect the topological signal, plotting ${\cal
R}(\Omega_0,\Omega_\Lambda)$.  \item If there is any topological
signal, the position of the spike gives the values of the cosmological
parameters on the scale of the catalog's limit (see figure
\ref{testR}).
\end{enumerate}

We proceed as follows.  We first generate a catalog by choosing the
number of objects in the fundamental domain ($A=30$), the topology
(Weeks manifold) and the cosmological parameters (e.g.
$\Omega_0=0.2$, $\Omega_\Lambda=0.1$), we then use a second code to
apply the test and we draw ${\cal R}$ in terms of the two cosmological
parameters.  As shown in \cite{uzan99}, the method works pretty well
in the sense that there is a strong spike which signals a non trivial
topology and which determines the cosmological parameters.  But we
also see that a slight deviation in the evaluation of the cosmological
parameters would make the spike to disappear.  This effect is now
discussed.

\subsection{\bf Real data}

When one wants to apply the CCP--method to real data, one has to face
a number of problems.

\begin{figure}
\centering
\epsfig{figure=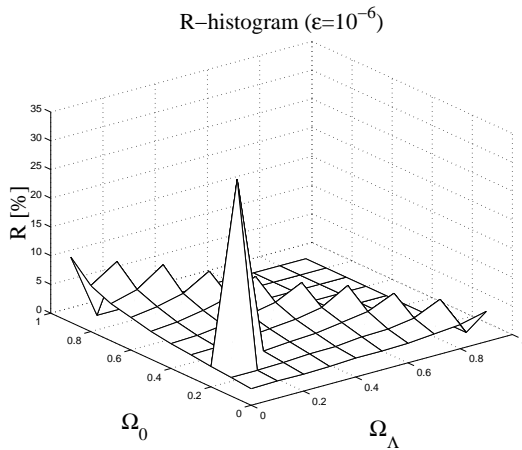, width=7cm} 
\caption{Computation of the 
CCP--index on a simultated catalog of depth $z=3$ in a universe model
with the Weeks topology, $\Omega_0=0.2$, $\Omega_\Lambda=0.1$, using a
bin resolution $\epsilon=10^{-6}$.}
\label{testR}
\end{figure}

First, we cannot use a zero width bin, one of the reasons being that
the sources are not comoving.  In \cite{uzan99}, we estimate the
precision needed on the cosmological parameters when working with a
bin width $\epsilon$ such that
\begin{equation}
\left|\frac{\delta\Omega_0}{\Omega_0}\right|\simeq\epsilon.
\end{equation}
Indeed, using a catalog with a smaller depth $z_{\rm{max}}=3$ will
allow us to use a smaller resolution for the cosmological parameters.
One has thus to find a compromise between depth and resolution as
discussed in \cite{uzan99}.  On figure \ref{testR}, we give an example
with a bin width of $\epsilon=10^{-6}$ which produces a background
noise.

Now, we apply our test to the quasars catalog \cite{quascat} used
ahead (see figure \ref{catq}).  No topological signature was found
\cite{uzan99}.  Does it mean that there is no topological effect on
scales smaller than $z_{\rm{max}}\simeq3$~?  Not necessarily, since we
applied the test with presicions $\epsilon=10^{-7}, 10^{-6}, 10^{-5}$
and were unable to span the full cosmological parameters' space with
the required accuracy.  The computational time is one of the main
limitations of our technique (this point is discussed extensively in
\cite{uzan99})

Now, another limitation comes from the peculiar velocities of the
sources.  In \cite{uzan99}, we discussed in details the effect of
peculiar velocities both on $xy$--pairs and on $xg(x)$--pairs, we and
showed that even if the amplitude of the topological signal is
reduced, it is not destroyed.

\section{CONCLUSIONS}

We have reviewed the main allowed topological structures for
relativistic universe models and we discussed the various methods
aimed to detect this topology.  Focusing on 3D-methods, we described
the crystallographic method and we applied it to a quasars catalog to
obtain a new constraint for Euclidean manifolds,
$L_0\geq3000\,h^{-1}\,\hbox{Mpc}$.  Then we discussed the failure of
this method for detecting the topology of compact hyperbolic
manifolds.  We thus generalised the method by introducing a new
technique based on the construction of a CCP--index.  The main
difference with the cosmic crystallography method is the collecting
process of all correlated pairs of the catalog, which enhances the
signal associated to the existence of a non trivial topology.  We gave
examples of computation of such an index in order to show its
statistical relevance.  We finally discussed the implementation
difficulties of this new method when working with observational data.

\end{document}